\documentclass[twocolumn,noshowpacs,noshowkeys,preprintnumbers,amsmath,amssymb,prb]{revtex4}
\usepackage{graphicx,tabularx}
\usepackage{amssymb}
\usepackage[mathcal]{euscript}
\begin{document}
\title{A model for the formation energies of alanates and boranates}
\author{Michiel J. van Setten}
\affiliation{Electronic Structure of Materials, Institute for Molecules and Materials, Faculty
of Science,\\ Radboud University Nijmegen, Toernooiveld 1, 6525 ED Nijmegen, The Netherlands}
\author{Geert Brocks}
\affiliation{Computational Materials Science, Faculty of Science and Technology and MESA+ Institute
for Nanotechnology, University of Twente, P.O. Box 217, 7500 AE Enschede, The Netherlands}
\author{Gilles A. de Wijs}
\affiliation{Electronic Structure of Materials, Institute for Molecules and Materials, Faculty
of Science,\\ Radboud University Nijmegen, Toernooiveld 1, 6525 ED Nijmegen, The Netherlands}

\date{\today}



\begin{abstract}
We develop a simple model for the formation energies (FEs) of alkali and alkaline earth alanates
and boranates, based upon ionic bonding between metal cations and AlH$_4^-$ or BH$_4^-$ anions. The
FEs agree well with values obtained from first principles calculations and with experimental FEs.
The model shows that details of the crystal structure are relatively unimportant. The small size of the BH$_4^-$ anion causes a strong bonding in the crystal, which makes boranates more stable than
alanates. Smaller alkali or alkaline earth cations do not give an increased FE. They involve a
larger ionization potential that compensates for the increased crystal bonding.
\end{abstract}

\maketitle

The large scale utilization of hydrogen as a fuel crucially depends on the development of compact
storage materials with a high mass content of hydrogen.\cite{bog_rev} Over the last decade alanates
and boranates have been studied extensively because of their potential use as hydrogen storage
materials.\cite{bog_rev,zuttel04} These materials consist of a lattice of metal cations and
AlH$_4^-$ or BH$_4^-$ anions, respectively. The ideal hydrogen storage material should have a high
gravimetric hydrogen density, which requires the use of light metals. Moreover, the formation
energy (FE) of such a material has to be such that it is stable at room temperature, yet it has to
decompose at low temperature to release its hydrogen. In principle a large variety of alanates and
boranates can be synthesized by changing the metal cations, which can be used to tune the formation
energy.\cite{nakamori06}

Since synthesis is a very time consuming effort, there is a need for a materials specific theory
with a predictive power for the FE. At present the state of the art is formed by first principles
calculations based upon density functional theory (DFT). Several papers have been dedicated to
trends in the DFT FEs of alanates and
boranates.\cite{nakamori06,chun,vajeeston04,lovvik05rev,vajeeston05} There exists a surprising
variety of crystal structures among these compounds. In DFT calculations the crystal structure with
the lowest energy has to be searched for each compound, and the cell parameters and the atomic
positions have to be optimized. This procedure also makes DFT calculations a very time consuming
effort. A simple theory would help to understand the trend in the FEs of alanates and boranates.

Our aim is to construct a simple model for the FEs at 0 K of alkali alanates and boranates
(MAH$_4$, M = Li, Na, K; A = Al, B) and of their alkaline earth counterparts (M$'$(AH$_4$)$_2$, M =
Mg, Ca), avoiding the use of the actual crystal structure. We assume that these compounds can be described by ionic bonding between M$^+$ or
M$'^{2+}$ cations and AH$_4^-$ anions. Our model for the FE, $\Delta E_f$, is based upon a
Born-Haber cycle,\cite{born19}
\begin{equation}
\Delta E_f = E_\mathrm{elem} + E_\mathrm{ions} + E_\mathrm{crys}. \label{eq:hof}
\end{equation}
Starting from bulk elemental solids and H$_2$ molecules, $E_\mathrm{elem}$ is the energy required
to atomize the solids and the molecules. The $E_\mathrm{elem}$ of the bulk solids are listed in
Table~\ref{param}. We use a value of 4.48 eV for the dissociation energy of H$_2$.\cite{hcp}

The second step is to create M$^+$, M$'^{2+}$ and AH$_4^-$ ions from the atoms, represented by the
energy $E_\mathrm{ions}$. The contribution to $E_\mathrm{ions}$ from the M$^+$ ions is simply the
first ionization potential (IP) and from the M$'^{2+}$ ions it is the sum of the first and second
IPs. The numbers $\Sigma IP$ are given in Table~\ref{param}. We calculate the
contribution to $E_\mathrm{ions}$ from the AH$_4^-$ anions as follows. First an electron is added
to an Al or B atom, which lowers the energy by the atomic electron affinity (EA). These atoms then
have four valence electrons that are used to form covalent bonds with four hydrogen atoms. Using
EAs of 0.44 eV and 0.28 eV for Al and B, and 2.91 eV and 3.45 eV for the Al-H and B-H bond
strengths,\cite{hcp} we calculate FEs of $-12.08$ eV and $-14.08$ eV for the AlH$_4^-$ and BH$_4^-$
anions.

The final step consists of constructing the crystal from the M$^+$ (or M$'^{2+}$) and AH$_4^-$
ions, which is represented by the energy $E_\mathrm{crys}$. We use a simple Born model for the
potential between cations and anions. It consists of an attractive Coulomb potential between point
charges at the centers of the ions plus a repulsive short-range potential $\propto r^{-\bar{n}}$,
where $\bar{n}$ is the average Born exponent. $E_\mathrm{crys}$ is then given by\cite{ssp16}
\begin{equation}
E_\mathrm{crys} = \frac{M_c Z_A Z_C e^2}{4\pi \varepsilon_0 r_0}\left(1-\frac{1}{\bar{n}}\right),
\label{eq:born}
\end{equation}
where $Z_A=-1$ and $Z_C=+1,+2$ are the valencies of the anions and the cations, respectively, and
$r_0$ is the shortest cation-anion distance in the lattice. $M_c$ represents the Madelung constant,
which depends upon the type of lattice.\cite{ssp16}

\begin{table}[!tbp]
\begin{ruledtabular}
\caption{\label{param} Ionic radii, $R_{\textrm{ion}}$ (\AA), summed ionization potentials, $\Sigma
IP$ (eV), dissociation energies for the elemental bulks, $E_{\textrm{dis}}$ (eV), and the Born
exponents, $n$. Values are taken from Refs.~\onlinecite{hcp} and ~\onlinecite{shannon76}}
\begin{tabular}{lrrrrrrrr}
           & Li   & Na   & K     & Mg    & Ca & B &  Al\\
\hline
$R_{\textrm{ion}}$  & 0.90 & 1.16 & 1.52 &  0.86 &  1.14 &  &\\
$\Sigma IP$ & 5.39 & 5.14 & 4.34 & 22.67 & 17.98 &  & \\
$E_{\textrm{dis}}$  & 1.64 & 1.08 & 0.93 & 1.48 &  1.81 & 5.81 & 3.38\\
$n$           & 5 & 7 & 9 & 7 & 9 & 7 & 9\\
\end{tabular}
\end{ruledtabular}
\end{table}

Note that in all compounds considered here the AlH$_4^-$ and BH$_4^-$ anions have a tetrahedral
geometry. However, in the Born model of Eq.~(\ref{eq:born}), we have approximated these tetrahedra
by spheres. Our motivation for this is that we are interested in a simple model of
$E_\mathrm{crys}$ without having to take into account the full details of the crystal structure.
The cation-anion distance then is the sum of the ionic radii of the cation and the anion,
$r_0=r_C+r_A$. Since the cations we consider are mostly octahedrally coordinated, we use standard
ionic radii $r_C$ of 6-fold coordinated alkaline and alkaline earth ions, see
Table~\ref{param}.\cite{shannon76} As the radius of the anions $r_A$ we use the Al-H and B-H bond
lengths, which are 1.62~\AA\ and 1.20~\AA, respectively. $r_A$ then roughly corresponds to the
average of the maximum radius of an AH$_4^-$ ion (A = Al, B) and the minimum radius, which is the
radius of the central atom A. The values obtained for the cation-anion distance $r_0$ then
correspond to the average of the cation-Al/B and cation-H distances in the crystal.

Avoiding the full details of the crystal structure also leads to using average values for the
Madelung constants $M_c$ in Eq.~(\ref{eq:born}). The alkali alanates and boranates have an AB type
lattice, where A is the alkali cation, and B is the boranate or alanate anion. The variation of the
Madelung constant over different AB lattices is relatively small, so we use an average value
$M_c=1.76$. The root mean square deviation (rms) averaged over all AB lattice types is 4\%. A
similar reasoning holds for the alkaline earth alanates and boranates. They have an AB$_2$ lattice,
whose average Madelung constant is $M_c=2.40$ with a rms deviation of 4\%.

\begin{figure}
\includegraphics[width=8.5cm]{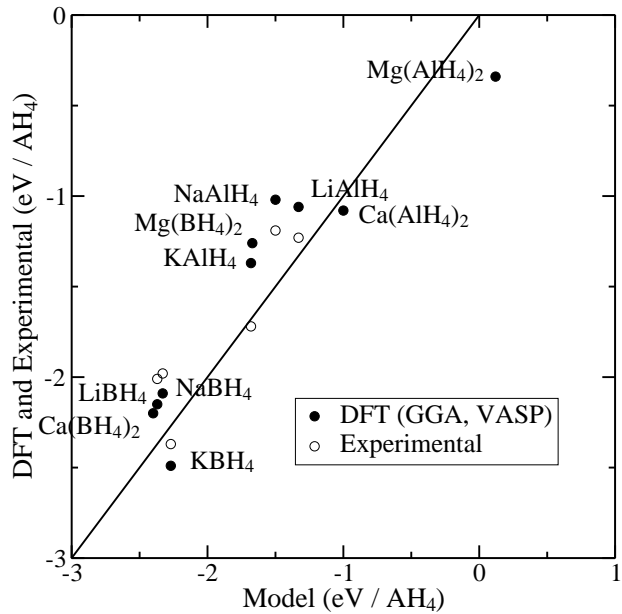}
\caption{\label{hofvaspmodexp} Model formation energies (eV/AH$_4$) compared to DFT and
experimental values.\cite{smith63}}
\end{figure}

Fig.~\ref{hofvaspmodexp} shows the most important results, i.e. the FE calculated with the model
represented by Eqs.~(\ref{eq:hof}) and (\ref{eq:born}), compared to experimental
values.\cite{smith63} For some of the materials the experimental FE is not known. Therefore we have
also performed first principles DFT calculations. We use the projector augmented wave (PAW)
method,\cite{paw,blo} and the PW91 generalized gradient approximation (GGA),\cite{pw91} as
implemented in the Vienna \em ab initio\em\ simulation package
(VASP).\cite{vasp1,vasp2,vasp3,vasp4} To integrate the Brillouin zone we apply the tetrahedron
scheme. The $\mathbf{k}$-point mesh and the plane wave kinetic energy cutoff (700 eV) are chosen
such, that total energies are converged to a numerical accuracy of 1~meV per formula
unit.\cite{zpe} The structures of LiAlH$_4$,\cite{vansetten06} NaAlH$_4$,\cite{vansetten06} Mg(AlH$_4$)$_2$,\cite{vansetten06}
Ca(AlH$_4$)$_2$,\cite{lovvik05} Ca(BH$_4$)$_2$,\cite{miwa:155122} KAlH$_4$,\cite{vajeeston04-2} and of the alkali
boranates\cite{vajeeston05} are taken from the literature. We additionally relaxed the atomic
positions, but the relaxations were small and had only a minor effect on the total energies. Our
calculated values compare well to those obtained in previous calculations.\cite{nakamori06,vajeeston05,lovvik05rev,chun,chou} Details on Mg(BH$_4$)$_2$ will be published elsewhere.\cite{vansetten06-2}

Fig.~\ref{hofvaspmodexp} shows that, despite its simplicity, the model gives FEs that are in
quantitative agreement with both the experimental, and the calculated DFT values. The rms deviation
of the model with the experimental and the DFT values is 0.27 and 0.33~eV/AH$_4$, respectively.
Note that these numbers are comparable to the rms deviation between the experimental and the first
principles values, 0.19~eV/AH$_4$, which represents the state-of-the-art. An obvious source of
error is our neglect of the details of the crystal structure, e.g., by using an average Madelung
constant in Eq.~(\ref{eq:born}). Changing the Madelung constant by 5\% changes the FE of the alkali
compounds by 0.4~eV/AH$_4$ and that of the alkaline earth compounds by 0.7~eV/AH$_4$. As these
numbers are larger than the rms deviation of the model, one can conclude that the details of the
crystal structure are relatively unimportant.

The model also seems to work reasonably well for some other boranates. The model FE for
Sc(BH$_4$)$_3$ is 0.37~eV/BH$_4$ higher than the DFT value calulated by Nakamori \em et.\ al\em,\cite{nakamori06} which is within the rms error bar given above. The model FE for
Zn(BH$_4$)$_2$ and CuBH$_4$ are 0.6 and 0.8~eV/BH$_4$ higher than the DFT values, respectively. For
cations with a nominal charge $Z_C=4$, such as Zr or Hf, the model breaks down. The model FE then
deviates by 2.5~eV/BH$_4$ from the DFT values.\cite{nakamori06} As can be seen from
Eq.~(\ref{eq:born}), for large $Z_C$ $E_\mathrm{crys}$ becomes sensitive to small changes in the
Madelung constant and the ionic radius of the cation, or in other words, to the details of the
crystal structure.

\begin{figure}
\centering
\includegraphics[angle=270,width=8.5cm]{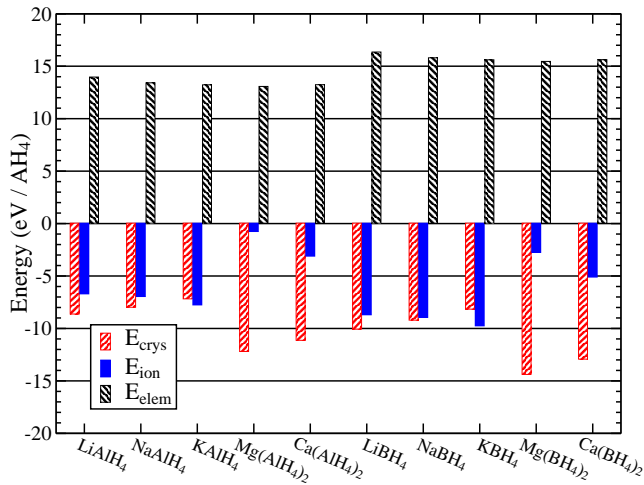}
\caption{\label{bar}The contributions to the formation energy (eV/AH$_4$) according to
Eq.~(\ref{eq:hof}).}
\end{figure}

Fig.~\ref{hofvaspmodexp} shows that boranates are generally more stable than alanates. The origin
of this stability can be analyzed by decomposing the FE into the contributions according to
Eq.~(\ref{eq:hof}), which is shown Fig.~\ref{bar}. The differences in formation energy of the
elements $E_\mathrm{elem}$ are to a large degree compensated by the differences in the formation
energies $E_\mathrm{ions}$ of the ions from the atoms. The ionic crystal energy $E_\mathrm{crys}$
of the boranates is however significantly larger than that of the alanates, which results in a
larger stability of the latter. This is a size effect since the BH$_4^-$ anions are significantly
smaller than the AlH$_4^-$ anions.

It has been observed that the dissociation energies of complex alkali hydrides into simple alkali
hydrides increase with the atomic number of the alkali atom.\cite{arr} For the FEs from the
elements the overall trend is not that clear. $E_\mathrm{elem}$ and $E_\mathrm{ions}$ both decrease
with increasing atomic number, see Fig.~\ref{bar}, which increases the stability. However, this is
almost compensated by $E_\mathrm{crys}$, which increases with the cation radius $r_C$. 

In the alkaline earth series the FE decreases with the atomic number. The dominant effect is a decreasing
$E_\mathrm{ions}$, which is due to a decrease in the ionization potentials of the cations, see
Table~\ref{param}.


To summarize we constructed a model for the formation energies (FEs) of alkali and alkaline earth
alanates and boranates from the elemental solids and H$_2$ molecules. The model is based upon ionic
bonding between metal cations and AlH$_4^-$ or BH$_4^-$ anions. It can be constructed using simple
energy values that are available in the literature and it does not make use of explicit crystal
structure information. Compared to experimental values, the model FEs have a similar accuracy as
calculated DFT values. The trends in the FEs over the series of compounds can be analyzed in terms
of the individual contributions to the model.

The authors wish to thank R.~A.~de~Groot for useful and stimulating discussions. This work is part
of the research programs of `Advanced Chemical Technologies for Sustainability (ACTS)' and the
`Stichting voor Fundamenteel Onderzoek der Materie FOM)', both financially supported by the
`Nederlandse Organisatie voor Wetenschappelijk Onderzoek (NWO)'.


\begin{thebibliography}{10}
\bibitem{zpe}
In both the DFT calculations and the model no zero point energies are taken
  into account.
\end{thebibliography}

\end{document}